\documentclass[12pt]{article}
\usepackage{amsmath,amssymb}

\newcommand{\eqdef}{\stackrel{\text{def}}{=}}
\newcommand{\n}{\nonumber \\}

\renewcommand{\theequation}{\arabic{section}.\arabic{equation}}

\newcommand{\maprightu}[2]
  {\smash{\mathop{\hbox to #1{\rightarrowfill}}\limits^{#2}}}
\newcommand{\maprightd}[2]
  {\smash{\mathop{\hbox to #1{\rightarrowfill}}\limits_{#2}}}

\newcommand{\mapdownr}[1]
  {\Bigg\downarrow\rlap{$\vcenter{\hbox{$\scriptstyle#1$}}$}}

\allowdisplaybreaks[4]

\begin{document}

\baselineskip=20pt

\newfont{\elevenmib}{cmmib10 scaled\magstep1}
\newcommand{\preprint}{
     \begin{flushleft}
       \elevenmib Yukawa\, Institute\, Kyoto\\
     \end{flushleft}\vspace{-1.3cm}
     \begin{flushright}\normalsize  \sf
       YITP-09-21\\
       March 2009
     \end{flushright}}
\newcommand{\Title}[1]{{\baselineskip=26pt
     \begin{center} \Large \bf #1 \\ \ \\ \end{center}}}
\newcommand{\Author}{\begin{center}
     \large \bf  Ryu Sasaki \end{center}}
\newcommand{\Address}{\begin{center}
       Yukawa Institute for Theoretical Physics,\\
       Kyoto University, Kyoto 606-8502, Japan
     \end{center}}
\newcommand{\Accepted}[1]{\begin{center}
     {\large \sf #1}\\ \vspace{1mm}{\small \sf Accepted for Publication}
     \end{center}}

\preprint
\thispagestyle{empty}
\bigskip\bigskip\bigskip

\Title{Exactly Solvable Birth and Death Processes}
\Author

\Address
\vspace{1cm}

\begin{abstract}
Many examples of exactly solvable birth and death processes, a typical stationary Markov chain, are presented together with the explicit expressions of the transition probabilities.
They are derived by similarity transforming exactly solvable `matrix' quantum mechanics, which is recently
proposed by Odake and the author. The ($q$-)Askey-scheme of hypergeometric orthogonal polynomials of a discrete variable and their dual polynomials play a central role.
The most generic solvable birth/death rates are rational functions of $q^x$ 
($x$ being the population) corresponding to the $q$-Racah polynomial.
\end{abstract}

\section{Introduction}
\label{intro}

The Brownian motion, a typical stationary Markov process with a continuous 
state space, is known to be described well by the Fokker-Planck
equation \cite{risken,feller}.
A {\em birth and death process\/}, on the other hand, being a typical stationary Markov
chain with a set of non-negative integers as a state space \cite{schoutens}, can be naturally considered as a discretisation of a one-dimensional Fokker-Planck equation.
Although birth and death processes have a wide range of applications \cite{feller,schoutens}, demography, queueing theory, inventory models and chemical dynamics, we will focus on their mathematical aspect, {\em i.e.,} the exact solvability.
In this paper we present 18 {\em exactly solvable\/} birth and death processes
based on the ($q$-)Askey scheme of hypergeometric orthogonal polynomials
having  discrete orthogonality measures.  They are also called orthogonal polynomials
of a discrete variable \cite{askey,ismail,nikiforov}.
For example, they are the ($q$-)Racah, the ($q$-)(dual)Hahn, the ($q$-)Krawtchouk,
the ($q$-)Charlier and the ($q$-)Meixner polynomials \cite{askey,ismail,koeswart}.
Various expressions of the transition probability are given explicitly together with the 
totality of the eigenvalues and the measures of the Karlin-McGregor type
representation \cite{KarMcG}.

It is well-known that  the one-dimensional Fokker-Planck
equation is related by a similarity transformation to a corresponding one-dimensional time-independent Schr\"odinger equation \cite{risken} or the eigenvalue problem for 
a suitable Hamiltonian. In other words, solutions of an exactly solvable Schr\"odinger equation
give the solutions of  the  corresponding Fokker-Planck equation, which is now 
exactly solvable.
The exact solvability means that the totality of the eigenvalues 
(in these cases, all are discrete) and the corresponding eigenfunctions are 
obtained exactly.
Here the Hamiltonian in quantum mechanics is an hermitian (self-adjoint)
linear operator in a certain Hilbert space. 
A natural {\em discretisation\/} of the Hamiltonians of 1-d quantum mechanics is 
hermitian matrices of a finite or infinite dimensions.
Recently exactly solvable `matrix' quantum mechanics was proposed by Odake and the present author \cite{os12}
by adopting special types of {\em tri-diagonal\/} Jacobi
matrices of finite or infinite dimensions as Hamiltonians.
The eigenfunctions are spanned by the above mentioned orthogonal
polynomials of a discrete variable. The corresponding discretisation of
the Fokker-Planck equation is, as expected,
the birth and death process with a {\em reflecting wall}(s)\eqref{BDeq}.
Among the 18 exactly solvable birth and death processes to be explored in this
paper, some are quite well-known having the linear 
\cite{feller, schoutens,KarMcGLin,KarMcGEhr} and quadratic \cite{vanaspralen} birth and death rates,
corresponding to the Meixner \S\ref{[KS1.9]}, Charlier \S\ref{[KS1.12]}, Krawtchouk \S\ref{[KS1.10]} and  Hahn \S\ref{[KS1.5]} polynomials.
The others have rational functions (of the population $x$) of the birth and death rates
corresponding to the dual Hahn \S\ref{[KS1.6]} and Racah \S\ref{[KS1.2]} polynomials
and some others have $q^{\pm x}$- linear, quadratic and rational  birth and death rates.
The most generic one is the $q$-Racah polynomial \S\ref{[KS3.2]} having $q^x$ rational birth and death
rates \eqref{qracahbd}.

This paper is organised as follows. In section two, the general properties 
of the Hamiltonians in 1-d quantum mechanics  (and/or the hermitian matrices) are reviewed in 
\S\ref{hamproperty}.
The relationship between the Schr\"odinger equation and the corresponding
Fokker-Planck equation is recapitulated in \S\ref{FPequationsect} and 
the solutions of the initial value problem of the Fokker-Planck equations and the
transition probabilities are expressed in terms of the orthogonal polynomials constituting the eigenfunctions of the corresponding Scr\"odinger equation.
In section three the birth and death operator is derived from the generic form of the
Hamiltonian of the exactly solvable `matrix' quantum mechanics of  \cite{os12}.
The solutions of the initial value problem of the birth and death  equations and the
transition probabilities are expressed in terms of the orthogonal polynomials constituting the eigenfunctions of the corresponding  Scr\"odinger equation of the `matrix' quantum mechanics. Various equivalent expressions of the transition probabilities are derived in terms of the dual polynomials.
Section four provides various data, the birth and death rates, the energy spectra and the sinusoidal coordinates, the stationary probability, the normalisation constants, and the eigenpolynomials,  of the exactly solvable 18 models, which are sufficient to evaluate the transition probability explicitly.
These 18 models  are named after the eigenpolynomials,
such as the  ($q$-)Racah, etc.  
The final section is for a brief summary and comments.
Appendix A provides the collection of the definitions of basic symbols
and functions for self-containedness. 
Throughout this paper we use the parameter $q$ in the range $0<q<1$.

\section{Fokker-Planck Operator from Hamiltonian}
\label{FPH}
\setcounter{equation}{0}

Here we recapitulate the well-known connection between the Fokker-Planck equation
and the Schr\"odinger equation \cite{risken} in order to introduce
appropriate notation and settings for the main purpose of the paper;
connecting the birth and death process to the `matrix' quantum mechanics
to be explored in the next section. 

\subsection{Properties of Hamiltonians}
\label{hamproperty}

Throughout this paper we discuss one degree of freedom systems  only.
The Hamiltonians to be discussed in this paper are {\em time independent\/}
and  share the  properties listed below. 
Most properties are common to the  the Hamiltonians having the continuous dynamical variable $x$  (to be used for the
Fokker-Planck equation) and the discrete dynamical variable $x$ (to be
applied to the birth and death processes). They are expressed by the same symbols. When they need different symbols, like the $L^2$ and $\ell^2$ norms, 
two different expressions are shown in a curly bracket as in \eqref{innerpro} and \eqref{inner2}. The upper (lower) one is for the continuous (discrete) dynamical variable
case. The former (the continuous variable) case corresponds to the ordinary quantum mechanics and the `discrete' quantum mechanics with the pure imaginary shifts \cite{os13}, which gives rise to the `deformed' Fokker-Planck equations \cite{hs1}.
\begin{description}
\item[(\romannumeral1)] {\em Factorisability\/}, 
\begin{equation}
\mathcal{H}=\mathcal{A}^\dagger\mathcal{A},
\label{factor1}
\end{equation}
in which ${}^\dagger$ denotes the  hermitian conjugation with respect to
the standard $L^2$ ($\ell^2$) inner product, see \eqref{innerpro}.
This also means that the Hamiltonian $\mathcal{H}$ is {\em positive semi-definite\/}.
\item[(\romannumeral2)] {\em Completeness  of its eigenfunctions\/} $\phi_n(x)$ {\em
belonging to discrete eigenvalues\/} (all distinct),
\begin{equation}
\mathcal{H}\phi_n(x)=\mathcal{E}(n)\phi_n(x),\quad
\mathcal{E}(0)<\mathcal{E}(1)<\cdots ,
\label{ham1}
\end{equation}
and all the eigenvectors are square normalisable and orthogonal with each other
\begin{equation}
(\phi_n,\phi_m)\eqdef \left\{
\begin{array}{c}
\int \phi_n(x)^*\phi_m(x)dx\\[6pt]
\sum_{x} \phi_n(x)^*\phi_m(x)
\end{array}
\right\}\
=\frac{1}{d_n^2}\delta_{n\,m},\quad 
0<d_n<\infty.
\label{innerpro}
\end{equation}

The range of the integration (summation) depends on the specific Hamiltonian.
Any element in the Hilbert space $\bf H$ is expanded by $\{\phi_n\}$:
\begin{equation}
\forall f\in{\bf H}\Rightarrow f=\sum_{n}f_n\hat{\phi}_n,\quad
\hat{\phi}_n\eqdef d_n\phi_n,\quad f_n\eqdef (\hat{\phi}_n,f).
\end{equation}
Here and hereafter $\hat{f}$ denotes a normalised vector 
$\hat{f}\eqdef f/\sqrt{(f,f)}$.
We choose all the eigenfunctions $\{\phi_n\}$ to be {\em real\/}, 
which is always possible in one-dimensional quantum mechanics.
\item[(\romannumeral3)]{\em The groundstate wavefunction\/} $\phi_0$ {\em is annihilated by\/}
$\mathcal{A}$ and is {\em positive everywhere\/},
\begin{equation}
\mathcal{A}\phi_0(x)=0 \quad \Rightarrow 
\mathcal{H}\phi_0(x)=0,\quad
\mathcal{E}(0)=0,\quad \phi_0(x)>0.
\label{Aphi0}
\end{equation}
\item[(\romannumeral4)] {\em The eigenfunction\/} $\phi_n(x)$ is $\phi_0(x)$ 
{\em times a polynomial\/},
\begin{equation}
\phi_n(x)=\phi_0(x)P_n(\eta(x)),\quad n=0,1,2,\ldots,
\quad P_0\equiv 1,
\label{eigenpoly}
\end{equation}
in which a real function $\eta(x)$ is called a {\em sinusoidal coordinate\/} 
\cite{os7,os12,os13}. In other words $P_n(\eta)$ is an orthogonal polynomial with 
the orthogonality measure $\phi_0(x)^2$
\begin{equation}
\left\{
\begin{array}{c}
\int\phi_0(x)^2P_n(\eta(x))P_m(\eta(x))dx\\[6pt]
\sum_{x}\phi_0(x)^2P_n(\eta(x))P_m(\eta(x))
\end{array}
\right\}\
=\frac{1}{d_n^2}\delta_{n\,m}.
\label{inner2}
\end{equation}
\item[(\romannumeral5)] {\em The similarity transformed Hamiltonian\/}
$\mathcal{H}$ with respect to $\phi_0(x)$ 
\begin{equation}
\widetilde{\mathcal{H}}\eqdef\phi_0^{-1}\circ \mathcal{H}\circ \phi_0
\label{tildeham}
\end{equation}

{\em provides a differential\/} ({\em difference\/})
{\em equation governing  the polynomial\/} $P_n(\eta(x))$.

\end{description}

\subsection{Fokker-Planck Equation}
\label{FPequationsect}

The Fokker-Planck equation in one dimension reads
\begin{equation}
\frac{\partial}{\partial t}\mathcal{P}(x;t)=L_{FP}\mathcal{P}(x;t),
\quad \mathcal{P}(x;t)\ge0,\quad \int \mathcal{P}(x;t)dx=1,
\label{FPeq}
\end{equation}
in which $\mathcal{P}(x;t)$ is the probability distribution over certain 
continuous range of the parameter $x$; for example $(-\infty,\infty)$, 
$(0,\infty)$ or $(0,\pi)$. The Fokker-Planck operator $L_{FP}$ 
corresponding to the Hamiltonian $\mathcal{H}$ \eqref{ham1}  is defined by \cite{risken,hs1}
\begin{equation}
L_{FP}\eqdef -\phi_0\circ\mathcal{H}\circ\phi_0^{-1},
\label{LFPdef}
\end{equation}
in which $\phi_0$ is defined in \eqref{Aphi0}.\footnote{
It should be emphasised that the {\em inverse similarity transformation\/}
in terms of $\phi_0$ is used here:
$
L_{FP}=-\phi_0^2\circ\widetilde{\mathcal{H}}\circ\phi_0^{-2}.
$
}
This guarantees that the eigenvalues of $L_{FP}$ are 
{\em negative semi-definite}.
The square normalised groundstate eigenfunction $\phi_0(x)$ provides
the {\em stationary distribution\/} $\hat{\phi}_0(x)^2$ of the corresponding Fokker-Planck
operator:
\begin{equation}
\frac{\partial}{\partial t}\hat{\phi}_0(x)^2=L_{FP}\hat{\phi}_0(x)^2=0,\quad
\int\hat{\phi}_0(x)^2dx=1.
\end{equation}
It is obvious that $\phi_0(x)\phi_n(x)$ is the eigenvector of the Fokker-Planck
operator $L_{FP}$:
\begin{equation}
L_{FP}\phi_0(x)\phi_n(x)=-\mathcal{E}(n)\phi_0(x)\phi_n(x),\quad
n=0,1,\ldots.
\end{equation}
Corresponding to an arbitrary initial probability distribution $\mathcal{P}(x;0)$,
(with $\int \mathcal{P}(x;0)dx=1$),
which can be expressed as a linear combination of 
$\{\hat{\phi}_0(x)\hat{\phi}_n(x)\}$, $n=0,1,\ldots$,
\begin{equation}
\mathcal{P}(x;0)=\hat{\phi}_0(x)\sum_{n=0}^{\infty}c_n\hat{\phi}_n(x),\quad
c_0=1,\quad
c_n\eqdef (\hat{\phi}_n,\hat{\phi}_0(x)^{-1}\mathcal{P}(x;0)),\quad n=1,2,\ldots,
\end{equation}
we obtain the solution of the Fokker-Planck equation
\begin{equation}
\mathcal{P}(x;t)=\hat{\phi}_0(x)\sum_{n=0}^{\infty}c_n\,e^{-\mathcal{E}(n)t}\hat{\phi}_n(x),
\quad
t>0.
\end{equation}
This is a consequence of the {\em completeness of the eigenfunctions\/} $\{\phi_n(x)\}$
(the polynomials) of the Hamiltonian $\mathcal{H}$.
The positivity of the spectrum $\mathcal{E}(n)>0$, $n\ge1$ \eqref{ham1} guarantees that 
the stationary distribution $\hat{\phi}_0^2(x)$ is achieved at future infinity:
\begin{equation}
\lim_{t\to\infty}\mathcal{P}(x;t)=\hat{\phi}_0^2(x).
\end{equation}

The {\em transition probability\/} from $y$ at $t=0$ 
(i.e., $\mathcal{P}(x;0)=\delta(x-y)$) to $x$ at $t$ 
is given by
\begin{equation}
\mathcal{P}(y,x;t)=\hat{\phi}_0(x)\hat{\phi}_0(y)^{-1}
\sum_{n=0}^{\infty}e^{-\mathcal{E}(n)t}\hat{\phi}_n(x)\hat{\phi}_n(y),
\quad
t>0.
\label{tranprobcont1}
\end{equation}
In terms of the polynomial $P_n(\eta(x))$, \eqref{eigenpoly}, it is expressed as
\begin{equation}
\mathcal{P}(y,x;t)=\phi_0(x)^2
\sum_{n=0}^{\infty}d_n^2\,e^{-\mathcal{E}(n)t}P_n(\eta(x))P_n(\eta(y)),
\quad
t>0,
\label{tranprobcont2}
\end{equation}
in which $d_n$ is the normalisation constant \eqref{innerpro},
\eqref{inner2}.

As shown in \cite{hs1} in some detail, various examples of exactly solvable quantum mechanics \cite{infhul,susyqm} and the `discrete' quantum mechanics with the pure imaginary shifts \cite{os13,os4,os7} provide many explicit cases in which the
transition probability \eqref{tranprobcont1}-\eqref{tranprobcont2} can be obtained exactly. The corresponding orthogonal polynomials are the Hermite, Laguerre and Jacobi polynomials in the ordinary quantum mechanics \cite{infhul,susyqm}
and the Meixner-Pollaczek, continuous (dual) Hahn, Wilson and Askey-Wilson polynomials \cite{hs1,os13} and their degenerate polynomials, like the continuous $q$-Hermite polynomials.

\section{Birth and Death process from `Matrix' Quantum Mechanics}
\label{BDgeneral}
\setcounter{equation}{0}

The birth and death equation  is a discretisation of the Fokker-Planck equation in one dimension \eqref{FPeq}. It reads
\begin{equation}
\frac{\partial}{\partial t}\mathcal{P}(x;t)=(L_{BD}\mathcal{P})(x;t),
\quad \mathcal{P}(x;t)\ge0,\quad \sum_x \mathcal{P}(x;t)=1,
\label{bdeqformal}
\end{equation}
in which $\mathcal{P}(x;t)$ is the probability distribution over a certain 
discrete set  of the parameter $x$. Here we  simply take a set of consecutive
non-negative integers, either finite or infinite:
\begin{equation}
x\in\mathbb{Z},\quad x\in[0,N]\text{  or  }
  [0,\infty).
\end{equation}
The {\em exactly solvable birth and death operator\/} or a matrix $L_{BD}$ is 
derived from the generic form of an exactly solvable Hamiltonian $\mathcal{H}$ of a `discrete' quantum mechanics with real shifts\:
\begin{equation}
  \mathcal{H}\eqdef
  -\sqrt{B(x)}\,e^{\partial}\sqrt{D(x)}
  -\sqrt{D(x)}\,e^{-\partial}\sqrt{B(x)}
  +B(x)+D(x),
  \label{genham}
\end{equation}
in which the two functions $B(x)$ and $D(x)$ are real and {\em positive}
but vanish at the boundary:
\begin{align}
  B(x)>0,\quad D(x)>0,\quad  D(0)=0\, ;\quad
  B(N)=0\ \ \text{for the finite case}.
  \label{BDcondition}
\end{align}
The explicit forms of the functions $B(x)$ and $D(x)$ are given in each subsection of section four, which are named after the orthogonal polynomials appearing as the main part of the
eigenfunctions.
In the Hamiltonian \eqref{genham} $e^{\pm\partial}$ are formal shift operators acting on a function $f$ of $x$ as
\begin{equation*}
(e^{\pm\partial}f)(x)=f(x\pm1).
\end{equation*}
Thus the Schr\"odinger equation   $\mathcal{H}\psi(x)=\mathcal{E}\psi(x)$ is a difference equation with real shifts:
\begin{align}
  \bigl(B(x)+D(x)\bigr)\psi(x)-\sqrt{B(x)D(x+1)}\,\psi(x+1)
  &-\sqrt{B(x-1)D(x)}\,\psi(x-1)
  =\mathcal{E}\psi(x),\n
  &x=0,1,\ldots,(N),\ldots.
  \label{diffeq}
\end{align}
The boundary condition $D(0)=0$ is necessary for the term $\psi(-1)$
 not to appear, and $B(N)=0$ is necessary  for the term $\psi(N+1)$
not to appear in the finite dimensional matrix case.

Although the Hamiltonian $\mathcal{H}$ \eqref{genham} is presented in
a difference operator form, it is in fact a real symmetric
{\em tri-diagonal\/} (Jacobi) matrix:
\begin{align}
  \mathcal{H}&=(\mathcal{H}_{x,y}),\qquad
  \mathcal{H}_{x,y}=\mathcal{H}_{y,x},\\
  \mathcal{H}_{x,y}&=
  -\sqrt{B(x)D(x+1)}\,\delta_{x+1,y}-\sqrt{B(x-1)D(x)}\,\delta_{x-1,y}
  +\bigl(B(x)+D(x)\bigr)\delta_{x,y}.
  \label{Jacobiform}
\end{align}
As mentioned above, the Hamiltonian is factorisable \eqref{factor1}, $\mathcal{H}=\mathcal{A}^\dagger\mathcal{A}$:
\begin{equation}
  \mathcal{A}^{\dagger}=\sqrt{B(x)}-\sqrt{D(x)}\,e^{-\partial},
  \qquad
  \mathcal{A}=\sqrt{B(x)}-e^{\partial}\sqrt{D(x)}.
  \label{A}
\end{equation}
In the matrix form
$\mathcal{A}^{\dagger}$ has the diagonal and sub-diagonal elements only
and $\mathcal{A}$ has  the diagonal and super-diagonal elements only
\begin{equation}
  (\mathcal{A}^{\dagger})_{x,y}=
  \sqrt{B(x)}\,\delta_{x,y}-\sqrt{D(x)}\,\delta_{x-1,y},\qquad
  \mathcal{A}_{x,y}=
  \sqrt{B(x)}\,\delta_{x,y}-\sqrt{D(x+1)}\,\delta_{x+1,y}.
\end{equation}
The equation \eqref{Aphi0} determining the groundstate wavefunction $\phi_0$ is easy to solve, since $\mathcal{A}\phi_0=0$ is a two term recurrence relation:
\begin{equation}
  \frac{\phi_0(x+1)}{\phi_0(x)}=\sqrt{\frac{B(x)}{D(x+1)}}
  \label{phi0/phi0=B/D}.
\end{equation}
It can be solved elementarily with the boundary (initial) condition   $\phi_0(0)=1$,
\begin{equation}
  \phi_0(x)=\sqrt{\prod_{y=0}^{x-1}\frac{B(y)}{D(y+1)}},\quad
  x=1,2,\ldots.
  \label{phi0=prodB/D}
\end{equation}
With the standard convention $\prod_{k=n}^{n-1}*=1$, the expression
\eqref{phi0=prodB/D} is valid for $x=0$, too.
For the infinite matrix case, the requirement of the finite
$\ell^2$ norm of the eigenvectors
\begin{equation}
  \sum_{x=0}^{\infty}\phi_0(x)^2=\sum_{x=0}^{\infty}
  \,\prod_{y=0}^{x-1}\frac{B(y)}{D(y+1)}<\infty
  \label{phizero2}
\end{equation}
imposes constraints on the asymptotic behaviours of $B(x)$ and $D(x)$.

With the above explicit form of the groundstate wavefunction $\phi_0(x)$, the similarity transformed Hamiltonian \eqref{tildeham} is easily obtained
\begin{equation}
  \widetilde{\mathcal{H}}\eqdef
  \phi_0^{-1}\circ \mathcal{H}\circ\phi_0
=B(x)(1-e^{\partial})+D(x)(1-e^{-\partial}).
\label{tildeham1}
\end{equation}
As mentioned above, $ \widetilde{\mathcal{H}}$ provides the difference equation for 
the polynomial eigenfunctions, 
\begin{equation}
(\widetilde{\mathcal{H}}P_n)(\eta(x))=\mathcal{E}(n)P_n(\eta(x)),
\end{equation}
that is,
\begin{equation}
B(x)\left(P_n(\eta(x))-P_n(\eta(x+1))\right)+
D(x)\left(P_n(\eta(x))-P_n(\eta(x-1))\right)=\mathcal{E}(n)P_n(\eta(x)).
\end{equation}
The eigenpolynomials $\{P_n\}$ are the orthogonal polynomials of a discrete variable.
See \S5 of \cite{os12} for various forms of $B(x)$ and $D(x)$ and the corresponding
orthogonal polynomails. It is also recapitulated in section 4 of this paper. For example,  they are the ($q$-)Racah, the ($q$-)(dual)Hahn, the ($q$-)Krawtchouk,
the ($q$-)Charlier and the ($q$-)Meixner polynomials \cite{askey,ismail,koeswart}.
As a  matrix,  $\widetilde{\mathcal{H}}$ is another tri-diagonal
matrix
\begin{equation}
  \widetilde{\mathcal{H}}=(\widetilde{\mathcal{H}}_{x,y}),\quad
  \widetilde{\mathcal{H}}_{x,y}=B(x)(\delta_{x,y}-\delta_{x+1,y})+
  D(x)(\delta_{x,y}-\delta_{x-1,y}).
\end{equation}

Corresponding to \eqref{LFPdef}, the {\em inverse similarity transformation\/} of the Hamiltonian $\mathcal{H}$ 
supplies the {\em birth and death operator\/} $L_{BD}$:
\begin{equation}
L_{BD}\eqdef -\phi_0\circ\mathcal{H}\circ\phi_0^{-1}
=(e^{-\partial}-1)B(x)+(e^{\partial}-1)D(x).
\label{LBDdef}
\end{equation}
Obviously the stationary probability is given by $\hat{\phi}_0(x)^2=d_0^2\phi_0(x)^2$.
In the matrix form, $L_{BD}$ is again  tri-diagonal:
\begin{equation}
  L_{BD}=({L_{BD}}_{x,y}),\quad
  {L_{BD}}_{x,y}=B(x-1)\delta_{x-1,y}-B(x)\delta_{x,y}+
  D(x+1)\delta_{x+1,y}-D(x)\delta_{x,y}.
  \label{LBDdefmat}
\end{equation}
In fact, $-L_{BD}$ is the transposed matrix of $\widetilde{\mathcal{H}}$:
\begin{equation}
-L_{BD}=(\widetilde{\mathcal{H}})^t,\quad -{L_{BD}}_{x,y}=\widetilde{\mathcal{H}}_{y,x}.
\end{equation}

With the explicit form of the birth and death operator $L_{BD}$, the {\em birth and death equation\/} \eqref{bdeqformal} in our notation reads
\begin{align}
\frac{\partial}{\partial t}\mathcal{P}(x;t)&=\sum_y{L_{BD}}_{x,y}\mathcal{P}(y;t)\n
&=-(B(x)+D(x))\mathcal{P}(x;t)+B(x-1)\mathcal{P}(x-1;t)+D(x+1)\mathcal{P}(x+1;t).
\label{BDeq}
\end{align}
The standard interpretation is that $x$ is the population of a group, 
$\mathcal{P}(x;t)$ is the probability for the group  to have the population $x$ at the time $t$,
and $B(x)$ is the  {\em birth rate\/},  $D(x)$ is the  {\em death rate\/}, respectively, when the population is $x$. It is quite easy to remember.
This is to be compared with the standard notation, for example, \cite{KarMcG}, XVII.5 of \cite{feller}, \S5.2 of \cite{ismail}:
\begin{equation}
\frac{\partial}{\partial t}p_n(t)=-(\lambda_n+\mu_n)p_n(t)+\lambda_{n-1}p_{n-1}(t)
+\mu_{n+1}p_{n+1}(t),
\end{equation}
in which $\lambda_n$  is the {\em birth rate\/} and $\mu_n$ is the {\em death rate\/}.
The following translation table of the notation will be helpful.
\begin{center}
 	\begin{tabular}{|c|c|c|}
    \hline
    	& standard \cite{feller,ismail}& this paper \\[4pt]
    \hline
    	population &$n=0,1,\ldots, (N), \ldots$ & $x=0,1,\ldots, (N), \ldots$ \\[4pt]
    \hline
    	probability&$p_n(t)$ & $\mathcal{P}(x;t)$ \\[4pt]
    \hline
    	Birth rate &$\lambda_n$\ \ ($\lambda_N=0$) & $B(x)$\ \ ($B(N)=0$) \\[4pt]
    \hline
	Death rate&$\mu_n$\ \ ($\mu_0=0$) & $D(x)$\ \ ($D(0)=0$) \\[4pt]
    \hline
 	\end{tabular}\\
\bigskip
	Table I: Translation Table.
\end{center}
The boundary condition for the finite case,  $\lambda_N=0$ ($B(N)=0$) \eqref{BDcondition} is said that the system  has a reflecting wall at the population 
$N$.

The {\em transition probability\/} from $y$ at $t=0$ 
(i.e., $\mathcal{P}(x;0)=\delta_{x,y}$) to $x$ at $t$ 
has exactly the same expression as that in the Fokker-Planck equation \eqref{tranprobcont1}
\begin{equation}
\mathcal{P}(y,x;t)=\hat{\phi}_0(x)\hat{\phi}_0(y)^{-1}
\sum_{n=0}e^{-\mathcal{E}(n)t}\hat{\phi}_n(x)\hat{\phi}_n(y),
\quad
t>0.
\label{tranprobdisc1}
\end{equation}
In terms of the polynomial $P_n(\eta(x))$, \eqref{eigenpoly}, it is expressed as
\begin{equation}
\mathcal{P}(y,x;t)=\phi_0(x)^2
\sum_{n=0}d_n^2\,e^{-\mathcal{E}(n)t}P_n(\eta(x))P_n(\eta(y)),
\quad
t>0.
\label{tranprobdisc2}
\end{equation}
It should be emphasised that in these formulas \eqref{tranprobdisc1}-\eqref{tranprobdisc2} everything is known including the measure  in contradistinction to the general formula by Karlin-McGregor \cite{KarMcG}.

Let us mention several equivalent expressions of the transition probability \eqref{tranprobdisc2} in terms of the {\em dual polynomials\/} \cite{leonard,bannaiito,terw,os12}. 
It is well-known that with proper normalisation
\begin{equation}
\eta(0)=0=\mathcal{E}(0),\quad P_0\equiv1\equiv Q_0,\quad P_n(0)=Q_x(0)=1,
\end{equation}
the two polynomials, $\{P_n(\eta)\}$ and
its {\em dual\/} polynomial $\{Q_x(\mathcal{E})\}$, coincide at
the integer lattice points \cite{os12}:
\begin{equation}
  P_n(\eta(x))=Q_x(\mathcal{E}(n)),\quad n=0,1,\ldots,(N),\ldots,
  \quad x=0,1,\ldots,(N),\ldots.
  \label{Duality}
\end{equation}
The dual polynomial $\{Q_x(\mathcal{E}(n))\}$, $x=0,1,\ldots$,  is a  {\em right eigenvector\/} of the similarity transformed Hamiltonian $\widetilde{\mathcal{H}}$ matrix with the eigenvalue 
$\mathcal{E}(n)$:
\begin{equation}
\sum_{y}\widetilde{\mathcal{H}}_{x,y}Q_y(\mathcal{E}(n))=\mathcal{E}(n)Q_x(\mathcal{E}(n)).
\end{equation}
The above equation is  the {\em three term recurrence relation\/} for the dual polynomials 
$\{Q_x(\mathcal{E})\}$:
\begin{align}
&\left(B(x)+D(x)\right)Q_x(\mathcal{E}(n))-B(x)Q_{x+1}(\mathcal{E}(n))
-D(x)Q_{x-1}(\mathcal{E}(n))=\mathcal{E}(n)Q_{x}(\mathcal{E}(n)),\\
&Q_0=1,\ Q_1(\mathcal{E})\!=\!(B(0)-\mathcal{E})/B(0),\
Q_2(\mathcal{E}\!)=\!(B(0)-\mathcal{E})(B(1)+D(1)-\mathcal{E})/(B(0)B(1)),\ldots.
\end{align}
For historical reasons, this polynomial $Q_x(\mathcal{E})$ is called the birth and death polynomial or the Karlin-McGregor polynomial \cite{KarMcG}.

In terms of the dual polynomials or the Karlin-McGregor polynomial, the transition probability is
\begin{equation}
\mathcal{P}(y,x;t)=\phi_0(x)^2
\sum_{n=0}d_n^2\,e^{-\mathcal{E}(n)t}Q_x(\mathcal{E}(n))Q_y(\mathcal{E}(n)),
\quad
t>0.
\label{tranprobdisc3}
\end{equation}
Following \cite{ismail}, let us introduce 
\begin{equation}
F_x(\mathcal{E}(n))\eqdef \phi_0(x)^2Q_x(\mathcal{E}(n)).
\end{equation}
Since $L_{BD}$ and $\widetilde{\mathcal{H}}$ is related by
\begin{equation}
L_{BD}=-\phi_0^2\circ \widetilde{\mathcal{H}}\circ \phi_0^{-2},
\end{equation}
it is easy to see that $F_x(\mathcal{E}(n))$ is a left eigenvector of $\widetilde{\mathcal{H}}$ and thus a right eigenvector of the birth and death operator $L_{BD}$:
\begin{align}
\sum_{y}{L_{BD}}_{x,y}F_y(\mathcal{E}(n))&=-\phi_0(x)^2\sum_{y}
\widetilde{\mathcal{H}}_{x,y}Q_y(\mathcal{E}(n))\n
&=-\mathcal{E}(n)\phi_0(x)^2Q_x(\mathcal{E}(n))
=-\mathcal{E}(n)F_x(\mathcal{E}(n)).
\end{align}
In terms of the right eigenvectors of $L_{BD}$, we obtain another expression of the transition probability \cite{ismail}
\begin{equation}
\mathcal{P}(y,x;t)=\frac{1}{\phi_0(y)^2}
\sum_{n=0}d_n^2\,e^{-\mathcal{E}(n)t}F_x(\mathcal{E}(n))F_y(\mathcal{E}(n)),
\quad
t>0.
\label{tranprobdisc4}
\end{equation}

The explicit forms of the transition probability \eqref{tranprobdisc1}, \eqref{tranprobdisc2}, \eqref{tranprobdisc3} and \eqref{tranprobdisc4} can be evaluated
straightforwardly if the Hamiltonian $\mathcal{H}$ of an exactly solvable discrete
quantum mechanics  is given. 
Thus we may call the functions $B(x)$ and $D(x)$ in  the Hamiltonian $\mathcal{H}$ of an exactly solvable discrete
quantum mechanics  \eqref{genham}, the birth and death rates of an {\em exactly solvable birth and death process\/}. As mentioned above, the association of the birth and death rates and
the orthogonal polynomial in this paper and in the literature \cite{KarMcG,ismail,vanaspralen} are dual to each other.
 Therefore the names of the polynomials in the next section
are the dual of the corresponding Karlin-McGregor polynomial except for the self-dual cases of the Krawtchouk \S\ref{[KS1.10]}, Meixner \S\ref{[KS1.9]} and Charlier \S\ref{[KS1.12]}.

In the subsequent section we will present 18 examples of exactly solvable birth and death processes.

\section{18 Examples}
\label{examples}
\setcounter{equation}{0}

Now let us proceed to give the 18 explicit examples of exactly solvable birth and death processes. The input is simply the function forms of the birth and death rates $B(x)$ and $D(x)$. The rest is calculable.  But here we also provide other data, taken from \cite{os12}, 
such as the energy eigenvalue $\mathcal{E}(n)$, the sinusoidal coordinate $\eta(x)$, 
the unnormalised stationary probability $\phi_0(x)^2$, the normalisation constants $d_n^2$ and the polynomials $P_n(\eta)$. 
Following the order of our previous work on the exactly solvable discrete quantum mechanics \cite{os12}, we  handle the most generic one first, and  then followed by the simpler ones. There is a logical reason for this order. The simpler ones are usually obtained by
specialising or restricting the parameters of the generic ones. Each example is 
called by the name of the corresponding orthogonal polynomial $P_n(\eta)$
 with the number {\em e.g.} [KS3.2] attached to it
indicating the subsection  in the standard review of Koekoek and Swarttouw \cite{koeswart}. 
 The finite ($N$) cases are discussed first and then the infinite ones. 
In each group the Askey-scheme of hypergeometric orthogonal
polynomials  (non-$q$ polynomials) will be discussed first and followed by the $q$-scheme polynomials.

Please note that the set of parameters is slightly different from the conventional ones \cite{askey,ismail,koeswart} for some polynomials, the reason explained  in \cite{os12}.
For some polynomials, for example, the ($q$-) Racah, (dual, $q$-) Hahn, etc, 
there are many non-equivalent parametrisations of $B(x)$ and $D(x)$, 
which could lead to non-equivalent birth and death processes.
Here we give only one of them as a representative, since the purpose of the paper is to show
exactly solvable structure, not to provide an exhaustive list of  all  solvable models.  See \cite{os12} for more general parametrisations and the allowed ranges of the parameters.
In the same spirit we did not include some of the polynomials listed in \cite{os12}.

\begin{center}
{\large\bf Finite Dimensional Cases}
\end{center} 

\subsection{Racah [KS1.2]}
\label{[KS1.2]}
The Racah polynomial is the most generic
hypergeometric orthogonal polynomial of a discrete variable.
All the other (non-$q$) polynomials are obtained by restriction
or limiting procedure. 
The function $B(x)$ and $D(x)$ depend on four real parameters $a$, $b$, $c$ and $d$,
with one of them, say $c$,  being related to $N$, $c\equiv -N$:
\begin{equation}
B(x)
  =-\frac{(x+a)(x+b)(x+c)(x+d)}{(2x+d)(2x+1+d)},\quad
  D(x)
  =-\frac{(x+d-a)(x+d-b)(x+d-c)x}{(2x-1+d)(2x+d)}.
\label{racahbd}
\end{equation}
The other data are:
\begin{align}
 \mathcal{E}(n)&= n(n+\tilde{d}),\quad
  \eta(x)=x(x+d),\quad 
  \tilde{d}\eqdef a+b+c-d-1,\\
  &\qquad\qquad\qquad\quad\  a\ge b,\  d>0,\ a>N+d,\ 0<b<1+d,
 \end{align}
 \vspace*{-9mm}
\begin{gather}  
  \phi_0(x)^2=\frac{(a,b,c,d)_x}{(1+d-a,1+d-b,1+d-c,1)_x}\,
  \frac{2x+d}{d},\\[4pt]
d_n^2=\frac{(a,b,c,\tilde{d})_n}
  {(1+\tilde{d}-a,1+\tilde{d}-b,1+\tilde{d}-c,1)_n}\,
  \frac{2n+\tilde{d}}{\tilde{d}}\times
  \frac{(-1)^N(1+d-a,1+d-b,1+d-c)_N}{(\tilde{d}+1)_N(d+1)_{2N}}.
\end{gather}
Here $(a)_n$ is the Pochhammer symbol \eqref{defPoch}.
Throughout this section, the format for $d_n^2$
consists of two parts separated by a $\times$ symbol:
$d_n^2=(d_n^2/d_0^2)\times d_0^2$.  The second part $d_0^2$ satisfies the
relation $\sum_x\phi_0(x)^2=1/d_0^2$.
The polynomial is 
\begin{align}
  &P_n(\eta(x))
  ={}_4F_3\Bigl(
  \genfrac{}{}{0pt}{}{-n,\,n+\tilde{d},\,-x,\,x+d}
  {a,\,b,\,c}\Bigm|1\Bigr),\
 \end{align}
in which ${}_4F_3$ is the standard hypergeometric series
\eqref{defhypergeom}. The dual polynomial is again the Racah polynomial with
the parameter correspondence
$(a,b,c,d)\leftrightarrow (a,b,c,\tilde{d})$.
The rational (a quartic polynomial  divided by a quadratic polynomial) birth and death rates
\eqref{racahbd} have not yet been discussed but the Racah polynomial appears in 
\cite{vanaspralen}.

\subsection{Hahn [KS1.5] }
\label{[KS1.5]}
This is a well-known example of quadratic (in $x$) birth and death rates
with two real positive parameters $a$ and $b$:
\begin{equation}
B(x)=(x+a)(N-x),\quad
  D(x)= x(b+N-x).
  \end{equation}
It has a quadratic energy spectrum
\begin{align}
\mathcal{E}(n)&= n(n+a+b-1),\quad
  \eta(x)=x,\quad
\phi_0(x)^2
 =\frac{N!}{x!\,(N-x)!}\,\frac{(a)_x\,(b)_{N-x}}{(b)_N},\\
  d_n^2
  &=\frac{N!}{n!\,(N-n)!}\,
  \frac{(a)_n\,(2n+a+b-1)(a+b)_N}{(b)_n\,(n+a+b-1)_{N+1}}
  \times\frac{(b)_N}{(a+b)_N},\\[4pt]
  P_n(\eta(x))
  &={}_3F_2\Bigl(
  \genfrac{}{}{0pt}{}{-n,\,n+a+b-1,\,-x}
  {a,\,-N}\Bigm|1\Bigr).
 \end{align}
The dual polynomial is the dual Hahn polynomial of the next subsection \ref{[KS1.6]}.
The quadratic birth and death rates are discussed in \cite{vanaspralen} associated with the dual Hahn polynomial.

\subsection{dual Hahn [KS1.6]}
\label{[KS1.6]}
The set of parameters is the same as the Hahn polynomial case.
The birth and death rates are rational functions of $x$,
\begin{equation}
B(x)=\frac{(x+a)(x+a+b-1)(N-x)}
  {(2x-1+a+b)(2x+a+b)},
\quad
  D(x)=\frac{x(x+b-1)(x+a+b+N-1)}
  {(2x-2+a+b)(2x-1+a+b)},
  \label{dualhahnBD2}
\end{equation}
giving rise to a linear energy spectrum
\begin{align}
&\mathcal{E}(n)=n,\quad
  \eta(x)= x(x+a+b-1),\quad
  \phi_0(x)^2
  =\frac{N!}{x!\,(N-x)!}
  \frac{(a)_x\,(2x+a+b-1)(a+b)_N}{(b)_x\,(x+a+b-1)_{N+1}},
  \label{dualhahneeta}\\
    &\qquad\qquad d_n^2
  =\frac{N!}{n!\,(N-n)!}\,\frac{(a)_n\,(b)_{N-n}}{(b)_N}
  \times\frac{(b)_{N}}{(a+b)_N},\\[4pt]
    &P_n(\eta(x))
  ={}_3F_2\Bigl(
  \genfrac{}{}{0pt}{}{-n,\,x+a+b-1,\,-x}
  {a,\,-N}\Bigm|1\Bigr). 
 \end{align}

\subsection{Krawtchouk [KS1.10] (self-dual)}
\label{[KS1.10]}
The case of linear birth and death rates are a very well-known example
(the Ehrenfest model) \cite{KarMcGEhr}
of an exactly solvable birth and death processes \cite{feller, schoutens}:
\begin{align}
  B(x)&=p(N-x),\quad
  D(x)=(1-p)x,\quad 0<p<1,\\
  \mathcal{E}(n)&=n,\qquad
  \eta(x)=x,\\
  \phi_0(x)^2&=
  \frac{N!}{x!\,(N-x)!}\Bigl(\frac{p}{1-p}\Bigr)^x,\quad
  d_n^2
  =\frac{N!}{n!\,(N-n)!}\Bigl(\frac{p}{1-p}\Bigr)^n\times(1-p)^N,\\[4pt]
    P_n(\eta(x))
  &={}_2F_1\Bigl(
  \genfrac{}{}{0pt}{}{-n,\,-x}{-N}\Bigm|p^{-1}\Bigr). 
\end{align}
This is a simplest example of self-dual polynomials. The stationary probability 
$\phi_0(x)^2d_0^2$ is the binomial distribution.

\subsection{$q$-Racah [KS3.2] }
\label{[KS3.2]}

This is the first example of the $q$-scheme of the orthogonal
polynomials. Among them the $q$-Racah polynomial is the most
generic. The set of parameters is  four real numbers $(a,b,c,d)$, 
which is different from the standard one in the same manner
as for the Racah polynomial. We restrict them
\begin{equation}
c=q^{-N},\ \ a\leq b,\ \ 0<d<1,\ \ 0<a<q^Nd,\  \ qd<b<1,\ \
\tilde{d}<q^{-1},\ \ \tilde{d}\eqdef abcd^{-1}q^{-1}.
\end{equation}
The functions $B(x)$ and $D(x)$ are
\begin{align}
 B(x)
  &=-\frac{(1-aq^x)(1-bq^x)(1-cq^x)(1-dq^x)}
  {(1-dq^{2x})(1-dq^{2x+1})}\,,\\[4pt]
  D(x)
 & =- \tilde{d}\,
  \frac{(1-a^{-1}dq^x)(1-b^{-1}dq^x)(1-c^{-1}dq^x)(1-q^x)}
  {(1-dq^{2x-1})(1-dq^{2x})}.
  \label{qracahbd}
\end{align}
The other data are
\begin{align}
  &\mathcal{E}(n)=(q^{-n}-1)(1-\tilde{d}q^n),\qquad
  \eta(x)=(q^{-x}-1)(1-dq^x),\\
 & \phi_0(x)^2=\frac{(a,b,c,d\,;q)_x}
  {(a^{-1}dq,b^{-1}dq,c^{-1}dq,q\,;q)_x\,\tilde{d}^x}\,
  \frac{1-dq^{2x}}{1-d},\\[4pt]
    &d_n^2
  =\frac{(a,b,c,\tilde{d}\,;q)_n}
  {(a^{-1}\tilde{d}q,b^{-1}\tilde{d}q,c^{-1}\tilde{d}q,q\,;q)_n\,d^n}\,
  \frac{1-\tilde{d}q^{2n}}{1-\tilde{d}}
  \times
  \frac{(-1)^N(a^{-1}dq,b^{-1}dq,c^{-1}dq\,;q)_N\,\tilde{d}^Nq^{\frac12N(N+1)}}
  {(\tilde{d}q\,;q)_N(dq\,;q)_{2N}},\\
  &P_n(\eta(x))
  ={}_4\phi_3\Bigl(
  \genfrac{}{}{0pt}{}{q^{-n},\,\tilde{d}q^n,\,q^{-x},\,dq^x}
  {a,\,b,\,c}\Bigm|q\,;q\Bigr), 
\end{align}
in which ${}_4\phi_3$ is the basic hypergeometric series
\eqref{defqhypergeom} and $(a;q)_n$ is the $q$-Pochhammer symbol \eqref{defqPoch}.
The dual $q$-Racah polynomial is again the $q$-Racah polynomial with
the parameter correspondence
$(a,b,c,d)\leftrightarrow (a,b,c,\tilde{d})$.

\subsection{$q$-Hahn [KS3.6]}
\label{[KS3.6]}
The $q$-Hahn polynomial has two positive parameters $a$ and $b$ and the birth and death rates are  quadratic polynomials in $q^x$:
\begin{equation}
B(x)=(1-aq^x)(q^{x-N}-1),\quad
  D(x)= aq^{-1}(1-q^x)(q^{x-N}-b),\quad 0<a,b<1.
\end{equation}
The other data are
\begin{align}
 \mathcal{E}(n)
  &=(q^{-n}-1)(1-abq^{n-1}),\qquad
  \eta(x)=q^{-x}-1,\\
    \phi_0(x)^2
  &=\frac{(q\,;q)_N}{(q\,;q)_x\,(q\,;q)_{N-x}}\,
  \frac{(a;q)_x\,(b\,;q)_{N-x}}{(b\,;q)_N\,a^x}\,,\\[4pt]
  d_n^2
  &=\frac{(q\,;q)_N}{(q\,;q)_n\,(q\,;q)_{N-n}}\,
  \frac{(a,abq^{-1};q)_n}{(abq^N,b\,;q)_n\,a^n}\,
  \frac{1-abq^{2n-1}}{1-abq^{-1}}
  \times\frac{(b\,;q)_N\,a^N}{(ab\,;q)_N},\\[4pt]
  P_n(\eta(x))
  &={}_3\phi_2\Bigl(
  \genfrac{}{}{0pt}{}{q^{-n},\,abq^{n-1},\,q^{-x}}
  {a,\,q^{-N}}\Bigm|q\,;q\Bigr).
\end{align}
Obviously the $q$-Hahn and dual $q$-Hahn are dual to each other.

\subsection{dual $q$-Hahn [KS3.7]}
\label{[KS3.7]}
For obvious reasons, we adopt the same parameters $(a,b)$ for
the $q$-Hahn and dual $q$-Hahn polynomials. The birth and death rates are rational functions of $q^x$:

\begin{align}
  B(x)&=
  \frac{(q^{x-N}-1)(1-aq^x)(1-abq^{x-1})}
  {(1-abq^{2x-1})(1-abq^{2x})},\qquad 0<a,b<1,\\[4pt]
  D(x)&=aq^{x-N-1}
  \frac{(1-q^x)(1-abq^{x+N-1})(1-bq^{x-1})}
  {(1-abq^{2x-2})(1-abq^{2x-1})},\\[4pt]
  \mathcal{E}(n)&=q^{-n}-1,\qquad
  \eta(x)=(q^{-x}-1)(1-abq^{x-1}),\\[4pt]
  \phi_0(x)^2
  &=\frac{(q\,;q)_N}{(q\,;q)_x\,(q\,;q)_{N-x}}\,
  \frac{(a,abq^{-1}\,;q)_x}{(abq^N,b\,;q)_x\,a^x}\,
  \frac{1-abq^{2x-1}}{1-abq^{-1}}\,,\\[4pt]
  d_n^2
  &=\frac{(q\,;q)_N}{(q\,;q)_n\,(q\,;q)_{N-n}}\,
  \frac{(a\,;q)_n(b\,;q)_{N-n}}{(b;q)_N\,a^n}
  \times\frac{(b\,;q)_N\,a^N}{(ab;q)_N}\,,\\[4pt]
   P_n(\eta(x))
  &={}_3\phi_2\Bigl(
  \genfrac{}{}{0pt}{}{q^{-n},\,abq^{x-1},\,q^{-x}}
  {a,\,q^{-N}}\Bigm|q\,;q\Bigr).
 \end{align}

\subsection{quantum $q$-Krawtchouk [KS3.14]}
\label{[KS3.14]}
This has one positive parameter $p>q^{-N}$. The birth and death rates are
quadratic polynomials in $q^x$:
\begin{align}
  B(x)&=p^{-1}q^x(q^{x-N}-1),\qquad
  D(x)=(1-q^x)(1-p^{-1}q^{x-N-1}),\\
  \mathcal{E}(n)&=1-q^n,\qquad
  \eta(x)=q^{-x}-1,\\
  \phi_0(x)^2
  &=\frac{(q\,;q)_N}{(q\,;q)_x(q\,;q)_{N-x}}\,
  \frac{p^{-x}q^{x(x-1-N)}}{(p^{-1}q^{-N}\,;q)_x}\,,\\[4pt]
  d_n^2
  &=\frac{(q\,;q)_N}{(q\,;q)_n(q\,;q)_{N-n}}\,
  \frac{p^{-n}q^{-Nn}}{(p^{-1}q^{-n}\,;q)_n}\,
  \times(p^{-1}q^{-N}\,;q)_N,\\[4pt]
  P_n(\eta(x))
  &={}_2\phi_1\Bigl(
  \genfrac{}{}{0pt}{}{q^{-n},\,q^{-x}}{q^{-N}}\Bigm|q\,;pq^{n+1}\Bigr).
\end{align}
\subsection{$q$-Krawtchouk [KS3.15]}
\label{[KS3.15]}
This has one positive parameter $p>0$ and the
birth and death rates are
linear in $q^x$:
\begin{align}
  B(x)&=q^{x-N}-1,\qquad
  D(x)=p(1-q^x),\\
  \mathcal{E}(n)&=(q^{-n}-1)(1+pq^n),\qquad
  \eta(x)=q^{-x}-1,\\
  \phi_0(x)^2&=\frac{(q\,;q)_N}{(q\,;q)_x(q\,;q)_{N-x}}\,
  p^{-x}q^{\frac12x(x-1)-xN},\\
  d_n^2
  &=\frac{(q\,;q)_N}{(q;q)_n(q;q)_{N-n}}\,
  \frac{(-p\,;q)_n}{(-pq^{N+1}\,;q)_n\,p^nq^{\frac12n(n+1)}}\,
  \frac{1+pq^{2n}}{1+p}
  \times\frac{p^{N}q^{\frac12N(N+1)}}{(-pq\,;q)_N},\\[4pt]
P_n(\eta(x))
  &={}_3\phi_2\Bigl(
  \genfrac{}{}{0pt}{}{q^{-n},\,q^{-x},\,-pq^n}{q^{-N},\,0}\Bigm|q\,;q\Bigr).
\end{align}

\subsection{affine $q$-Krawtchouk [KS3.16] (self-dual)}
\label{[KS3.16]}
This has one positive parameter $p$ and the birth and death rates are
quadratic polynomials in $q^x$:
\begin{align}
  B(x)&=(q^{x-N}-1)(1-pq^{x+1}),\quad
  D(x)=pq^{x-N}(1-q^x),\quad 0<p<q^{-1}, \\
  \mathcal{E}(n)&=q^{-n}-1,\qquad
  \eta(x)=q^{-x}-1,\\
  \phi_0(x)^2&=\frac{(q\,;q)_N}{(q\,;q)_x(q\,;q)_{N-x}}\,
  \frac{(pq\,;q)_x}{(pq)^x}\,,\quad
  d_n^2
  =\frac{(q\,;q)_N}{(q\,;q)_n(q\,;q)_{N-n}}\,
  \frac{(pq\,;q)_n}{(pq)^n}\times(pq)^N,\\[4pt]
  P_n(\eta(x))
  &={}_3\phi_2\Bigl(
  \genfrac{}{}{0pt}{}{q^{-n},\,q^{-x},\,0}{pq,\,q^{-N}}\Bigm|q\,;q\Bigr).
\end{align}

\bigskip

\begin{center}
{\large\bf Infinite Dimensional Cases}
\end{center}
In contrast to the finite dimensional case, the structure of the
polynomials is severely constrained by the asymptotic forms of
the functions $B(x)$ and $D(x)$ \eqref{phizero2}.

\subsection{Meixner [KS1.9] (self-dual)}
\label{[KS1.9]}

This is the best known example of exactly solvable birth and death processes 
\cite{KarMcGLin} and
 the birth and death rates are both linear in $x$  with
simple linear energy spectra
$\mathcal{E}(n)=n$ and $\eta(x)=x$.
It has two positive parameters $\beta$ and $c$:
\begin{align}
B(x)&=\frac{c}{1-c}(x+\beta),\quad
  D(x)=\frac{1}{1-c}\,x,\quad \beta>0,\quad 0<c<1,
  \label{MeixnerBD}\\
\mathcal{E}(n)&=n,\qquad
  \eta(x)=x,
  \label{MeixnerEeta}\\
\phi_0(x)^2&=\frac{(\beta)_x\,c^x}{x!}\,,\quad
  d_n^2
  =\frac{(\beta)_n\,c^n}{n!}\times(1-c)^{\beta},
  \label{Meixnerphi0d}\\
  P_n(\eta(x))
  &={}_2F_1\Bigl(
  \genfrac{}{}{0pt}{}{-n,\,-x}{\beta}\Bigm|1-c^{-1}\Bigr).
  \label{MeixnerP}
\end{align}

\subsection{Charlier [KS1.12] (self-dual)}
\label{[KS1.12]}
This is another best known example of exactly solvable birth and death processes with 
a constant birth rates $a>0$ and a linear death rates:
\begin{align}
  B(x)&=a,\qquad
  D(x)=x,
  \label{charlBD}\\
  \mathcal{E}(n)&=n,\qquad
  \eta(x)=x,
  \label{charlEeta}\\
  \phi_0(x)^2&=\frac{a^x}{x!}\,,\qquad
  d_n^2
  =\frac{a^{n}}{n!}\times e^{-a},
  \label{charlphi0d}\\
  P_n(\eta(x)
  &={}_2F_0\Bigl(
  \genfrac{}{}{0pt}{}{-n,\,-x}{-}\Bigm|-a^{-1}\Bigr).
  \label{charlP}
\end{align}
The stationary probability $\phi_0(x)^2d_0^2$ \eqref{charlphi0d} is the Poisson distribution.

\subsection{little $q$-Jacobi [KS3.12]}
\label{[KS3.12]}
This has two parameters $a$ and $b$. The birth and death rates grow exponentially as $x$ tends to infinity:
\begin{align}
  B(x)&=a(q^{-x}-bq),\quad
  D(x)=q^{-x}-1,\quad 0<a<q^{-1},\quad b<q^{-1},\\
  \mathcal{E}(n)&=(q^{-n}-1)(1-abq^{n+1}),\qquad
  \eta(x)=1-q^x,\\
  \phi_0(x)^2&=\frac{(bq\,;q)_x}{(q\,;q)_x}(aq)^x,
  \label{littleqjacobiphi0}\\
  d_n^2
  &=\frac{(bq,abq\,;q)_n\,a^nq^{n^2}}{(q,aq\,;q)_n}\,
  \frac{1-abq^{2n+1}}{1-abq}
  \times\frac{(aq\,;q)_{\infty}}{(abq^2\,;q)_{\infty}}\,,
  \label{littleqjacobidn}\\[4pt]
  P_n(\eta(x))
  &=(-a)^{-n}q^{-\frac12n(n+1)}\frac{(aq\,;q)_n}{(bq\,;q)_n}\,
  {}_2\phi_1\Bigl(
  \genfrac{}{}{0pt}{}{q^{-n},\,abq^{n+1}}{aq}\Bigm|q\,;q^{x+1}\Bigr). 
   \label{littleqjacobinorm}
\end{align}
The normalisation of the polynomial is different from the conventional one.

\subsection{$q$-Meixner [KS3.13]}
\label{[KS3.13]}

This has two positive parameters $b$ and $c$. The birth and death rates are
quadratic in $q^x$ and as $x$ goes to infinity, 
the birth rates tend to zero and the death rates tend to unity:
\begin{align}
  B(x)&=cq^x(1-bq^{x+1}),\quad
  D(x)=(1-q^x)(1+bcq^x),\quad 0<b<q^{-1},\quad c>0,\\
  \mathcal{E}(n)&=1-q^n,\qquad
  \eta(x)=q^{-x}-1,\\
\phi_0(x)^2&=
  \frac{(bq\,;q)_x}{(q,-bcq\,;q)_x}\,c^xq^{\frac12x(x-1)},\quad
  d_n^2
  =\frac{(bq\,;q)_n}{(q,-c^{-1}q\,;q)_n}
  \times\frac{(-bcq\,;q)_{\infty}}{(-c\,;q)_{\infty}}\,,\\[4pt]
  P_n(\eta(x))
  &={}_2\phi_1\Bigl(
  \genfrac{}{}{0pt}{}{q^{-n},\,q^{-x}}{bq}\Bigm|q\,;-c^{-1}q^{n+1}\Bigr).
\end{align}

\subsection{little $q$-Laguerre/Wall [KS3.20] }
\label{[KS3.20]}
This has one positive parameter $a$ and both the birth and death rates grow 
exponentially as $x$ tends to infinity:
\begin{align}
  B(x)&=aq^{-x},\qquad\quad
  D(x)=q^{-x}-1,\qquad 0<a<q^{-1},\\
  \mathcal{E}(n)&=q^{-n}-1,\qquad
  \eta(x)=1-q^x,\\
  \phi_0(x)^2&=\frac{(aq)^x}{(q\,;q)_x}\,,\qquad
  d_n^2
  =\frac{a^nq^{n^2}}{(q,aq\,;q)_n}\times(aq\,;q)_{\infty}\,,\\[4pt]
  P_n(\eta(x))
  &={}_2\phi_0\Bigl(
  \genfrac{}{}{0pt}{}{q^{-n},\,q^{-x}}{-}\Bigm|q\,;a^{-1}q^x\Bigr).
  \label{littleqlaguerrenorm}
\end{align}
The normalisation of the polynomial is different from the conventional one.

\subsection{Al-Salam-Carlitz II [KS3.25] }
\label{[KS3.25]}
This has one positive parameter $a$ and the birth and death rates are 
quadratic in $q^x$. As $x$ goes to infinity the birth rates tend to zero and death rates
tend to unity:
\begin{align}
B(x)&=aq^{2x+1},\qquad\quad
  D(x)=(1-q^x)(1-aq^x),\quad 0<a<q^{-1},\\
  \mathcal{E}(n)&=1-q^n,\qquad\quad
  \eta(x)=q^{-x}-1,\\
  \phi_0(x)^2&=\frac{a^xq^{x^2}}{(q,aq\,;q)_x}\,,\quad
  d_n^2
  =\frac{(aq)^n}{(q\,;q)_n}\times(aq\,;q)_{\infty}\,,\\[4pt]
  P_n(\eta(x))
  &={}_2\phi_0\Bigl(
  \genfrac{}{}{0pt}{}{q^{-n},\,q^{-x}}{-}\Bigm|q\,;a^{-1}q^n\Bigr).
  \label{alsalamIInorm}
\end{align}
The normalisation of the polynomial is different from the conventional one.

\subsection{alternative $q$-Charlier [KS3.22]}
\label{[KS3.22]}

This has one positive parameter $a$. The birth rates are constant $a$ whereas the death rates grow exponentially as $x$ goes to infinity:
\begin{align}
  B(x)&=a,\quad
  D(x)=q^{-x}-1,\quad a>0,\\
  \mathcal{E}(n)&=(q^{-n}-1)(1+aq^n),\quad
  \eta(x)=1-q^x,\\
  \phi_0(x)^2&=\frac{a^xq^{\frac12x(x+1)}}{(q\,;q)_x}\,,
  \ \,
  d_n^2
  =\frac{a^nq^{\frac12n(3n-1)}}{(q\,;q)_n}\,
  \frac{(-a\,;q)_{\infty}}{(-aq^n\,;q)_{\infty}}\,
  \frac{1+aq^{2n}}{1+a}
  \times\frac{1}{(-aq\,;q)_{\infty}}\,,\\[4pt]
  P_n(\eta(x))
 & =q^{nx}\,{}_2\phi_1\Bigl(
  \genfrac{}{}{0pt}{}{q^{-n},\,q^{-x}}{0}\Bigm|q\,;-a^{-1}q^{-n+1}\Bigr).
  \label{alcharliernorm}
\end{align}
The normalisation of the polynomial is different from the conventional one.

\subsection{$q$-Charlier [KS3.23]}
\label{[KS3.23]}
This has one positive parameter $a$ and as $x$ goes to infinity the birth rates tend
to zero and the death rates tend to unity:
\begin{align}
  B(x)&=aq^x,\qquad\qquad
  D(x)=1-q^x,\quad a>0,\\
  \mathcal{E}(n)&=1-q^n,\qquad\quad
  \eta(x)=q^{-x}-1,\\
  \phi_0(x)^2&=\frac{a^xq^{\frac12x(x-1)}}{(q\,;q)_x}\,,\quad
  d_n^2
  =\frac{q^n}{(-a^{-1}q,q\,;q)_n}\times\frac{1}{(-a\,;q)_{\infty}}\,,\\[4pt]
  P_n(\eta(x))
  &={}_2\phi_1\Bigl(
  \genfrac{}{}{0pt}{}{q^{-n},\,q^{-x}}{0}\Bigm|q\,;-a^{-1}q^{n+1}\Bigr).
\end{align}

\section{Summary and Comments}
\label{summary}
\setcounter{equation}{0}

Following the simple line of arguments summarised in the following diagram, we presented
18 models of exactly solvable birth and death processes and their solutions, the transition probabilities.
In the diagram `ES' stands for Exactly Solvable.
\medskip
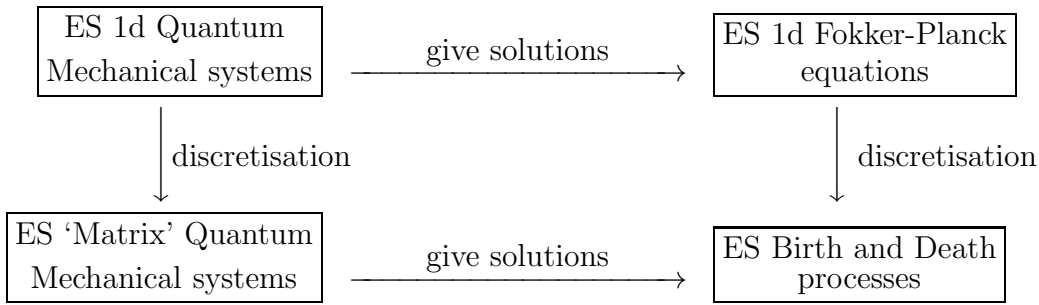
\begin{eqnarray*}
  \framebox{\shortstack{$\,$ES 1d Quantum $\,$ \\ \\Mechanical systems}}
  &\maprightu{45mm}{\mbox{give solutions}}&
  \framebox{\shortstack{ES 1d  Fokker-Planck\\ \\
     equations}}\\
  \mapdownr{\mbox{discretisation}}\hspace{20mm}&&
  \hspace{15mm}\mapdownr{\makebox{
     \shortstack{discretisation}}}\\
  \hspace*{-5mm}
  \framebox{\shortstack{ES  `Matrix' Quantum  \\ \\ Mechanical systems}}
  &\maprightu{45mm}{\mbox{give solutions}}&
  \framebox{\shortstack{ES Birth and Death  \\ \\
           processes}}
\end{eqnarray*}
The exactly solvable `matrix' quantum mechanics, or the 1-d `discrete' quantum mechanics
with real shifts was explored in detail in \cite{os12} to cover most of the hypergeometric
orthogonal polynomials of a discrete variable in the ($q$-) Askey scheme \cite{askey,ismail,koeswart}. For the `explanation' of the exact solvability, 
see a recent work  \cite{os14}. By comparing the present simple results with those in  the literature \cite{KarMcG, schoutens, ismail,vanaspralen} one would realise the essential role played by the
energy spectrum $\mathcal{E}(n)$ and the sinusoidal coordinate $\eta(x)$.
They are the eigenvalues of the two operators, called the Leonard pair, which characterise
the orthogonal polynomials completely \cite{leonard,bannaiito,terw}.

In this paper we did not discuss the generalisation of the birth and death processes  which  
has $\mu_0>0$ ($D(0)>0$), the {\em non-vanishing death rate at zero population\/}, although this has led to a new type of orthogonal polynomials in the cases when the 
birth and death rates $B(x)$ and $D(x)$ are linear and quadratic in $x$,
\cite{new1,new2}. It would be interesting to try further generalisation in this direction 
for which $B(x)$ and $D(x)$ are
rational, {\em e.g.\/} the Racah case \S\ref{[KS1.2]} or $q$-linear,  {\em e.g.\/} the $q$-Krawtchouk \S\ref{[KS3.15]}, or $q$-quadratic, {\em e.g.\/} the the affine $q$-Krawtchouk \S\ref{[KS3.16]},  or even the $q$-rational, {\em e.g.\/} the $q$-Racah \S\ref{[KS3.2]} cases.

It is a big challenge to try and find a closed form expression for
\begin{equation}
\sum_{n=0}d_n^2\,e^{-\mathcal{E}(n)t}P_n(\eta(x))P_n(\eta(y)),
\end{equation}
appearing as a part of the transition probability \eqref{tranprobcont2}, \eqref{tranprobdisc2} for various examples in section four. To the best of our knowledge, such expressions are known only for
the linear energy spectrum $\mathcal{E}(n)\propto n$. For example, for the Fokker-Planck equation
corresponding to the harmonic oscillator Hamiltonian, or the Ornshtein-Uhlenbeck process \cite{risken, hs1}, 
we have:
\begin{align}
\mathcal{H}&\eqdef-\frac{d^2}{dx^2}+x^2-1,\quad
\quad
L_{FP}=\frac{d^2}{dx^2}+2\frac{d}{dx} x,\quad \mathcal{E}(n)=2n,\quad \eta(x)=x,\\[4pt]
\mathcal{P}(y,x;t)&=\frac{e^{-x^2}}{\sqrt{\pi}}
\sum_{n=0}^\infty\frac{H_n(x)H_n(y)}{2^nn!}\,e^{-2nt}
=\frac{1}{\sqrt{\pi}\sqrt{1-e^{-4t}}}
\exp\left[-\frac{(x-y\,e^{-2t})^2}{1-e^{-4t}}\right].
\end{align}
The last equality was derived based on (6.1.13) of \cite{askey}. 
Another example is
\begin{align}
  \mathcal{H}&\eqdef-\frac{d^2}{dx^2}+x^2+\frac{g(g-1)}{x^2}-(1+2g),\quad
 L_{FP}=\frac{d^2}{dx^2}+2\frac{d}{dx}(x-\frac{g}{x}),\\
 & \qquad\qquad \mathcal{E}_n=4n,\quad \eta(x)=x^2, \quad \beta\eqdef g-1/2,\\
 \mathcal{P}(y,x;t)&=2e^{-x^2}x^{2g}
\sum_{n=0}^\infty\frac{n!\,L_n^{(\beta)}(x^2)L_n^{(\beta)}(y^2)}{\Gamma(n+\beta+1)}\,e^{-4nt}\\
&=\frac{2x^{2g}}{(1-e^{-4t})}
\exp\left[-\frac{(x^2+y^2\,e^{-4t})}{(1-e^{-4t})}\right](xye^{-2t})^{-\beta}I_{\beta}\left(\frac{2xye^{-2t}}{1-e^{-4t}}\right),
\end{align}
in which $I_\beta$ is the modified Bessel function of order $\beta$. 
The last equality was derived based on (6.2.25) of \cite{askey}.
We would like to ask experts in special functions and orthogonal polynomials to derive such 
bilinear generating functions for various energy spectra:
\begin{align}
\mathcal{E}(n)= n(n+d), \ q^{-n}-1,\ 1-q^n,\ (q^{-n}-1)(1-{d}q^n).
\end{align}
\section*{Acknowledgements}

We thank Mourad Ismail and Choon-Lin Ho who induced us to the present research.
This work is supported in part by Grants-in-Aid for Scientific
Research from the Ministry of Education, Culture, Sports, Science
and Technology, No.18340061 and No.19540179.

\section*{Appendix A: Some definitions related to the hypergeometric
and $q$-hypergeometric functions}
\label{appendA}
\setcounter{equation}{0}
\renewcommand{\theequation}{A.\arabic{equation}}

For self-containedness we collect several definitions related to
the ($q$-)hypergeometric functions \cite{koeswart}.

\noindent
$\circ$ Pochhammer symbol $(a)_n$ :
\begin{equation}
  (a)_n\eqdef\prod_{k=1}^n(a+k-1)=a(a+1)\cdots(a+n-1)
  =\frac{\Gamma(a+n)}{\Gamma(a)}.
  \label{defPoch}
\end{equation}
$\circ$ $q$-Pochhammer symbol $(a\,;q)_n$ :
\begin{equation}
  (a\,;q)_n\eqdef\prod_{k=1}^n(1-aq^{k-1})=(1-a)(1-aq)\cdots(1-aq^{n-1}).
  \label{defqPoch}
\end{equation}
$\circ$ hypergeometric series ${}_rF_s$ :
\begin{equation}
  {}_rF_s\Bigl(\genfrac{}{}{0pt}{}{a_1,\,\cdots,a_r}{b_1,\,\cdots,b_s}
  \Bigm|z\Bigr)
  \eqdef\sum_{n=0}^{\infty}
  \frac{(a_1,\,\cdots,a_r)_n}{(b_1,\,\cdots,b_s)_n}\frac{z^n}{n!}\,,
  \label{defhypergeom}
\end{equation}
where $(a_1,\,\cdots,a_r)_n\eqdef\prod_{j=1}^r(a_j)_n
=(a_1)_n\cdots(a_r)_n$.\\
$\circ$ $q$-hypergeometric series (the basic hypergeometric series)
${}_r\phi_s$ :
\begin{equation}
  {}_r\phi_s\Bigl(
  \genfrac{}{}{0pt}{}{a_1,\,\cdots,a_r}{b_1,\,\cdots,b_s}
  \Bigm|q\,;z\Bigr)
  \eqdef\sum_{n=0}^{\infty}
  \frac{(a_1,\,\cdots,a_r\,;q)_n}{(b_1,\,\cdots,b_s\,;q)_n}
  (-1)^{(1+s-r)n}q^{(1+s-r)n(n-1)/2}\frac{z^n}{(q\,;q)_n}\,,
  \label{defqhypergeom}
\end{equation}
where $(a_1,\,\cdots,a_r\,;q)_n\eqdef\prod_{j=1}^r(a_j\,;q)_n
=(a_1\,;q)_n\cdots(a_r\,;q)_n$.


\end{document}